\documentclass[aps,prc]{revtex4}
\usepackage{epsfig}

\newcommand{\be}{\begin{equation}}
\newcommand{\ee}{\end{equation}}
\newcommand{\bea}{\begin{eqnarray}}
\newcommand{\eea}{\end{eqnarray}}
\newcommand{\bean}{\begin{eqnarray*}}
\newcommand{\eean}{\end{eqnarray*}}

\newcommand{\gapproxeq}{\lower
.7ex\hbox{$\;\stackrel{\textstyle >}{\sim}\;$}}
\newcommand{\lapproxeq}{\lower
.7ex\hbox{$\;\stackrel{\textstyle <}{\sim}\;$}}

\newcommand{\tenbar}{\mbox{$\mathrm{\overline{\bf 10}}$} }
\def\3bar{$\bar {\hbox{\bf 3}}$}

\begin{document}

\title{\bf The Pentaquark $\Theta^+$ in $nK^+$ and $pK^0_S$, and its
 $\Sigma^+_5$ partner}

\author{
Qiang Zhao$^1$\footnote{e-mail: Qiang.Zhao@surrey.ac.uk}
and Frank E. Close$^2$\footnote{e-mail: F.Close1@physics.ox.ac.uk}
}
\affiliation{1) Department of Physics,
University of Surrey, Guildford, GU2 7XH, United Kingdom}
\affiliation{2) Department of Theoretical Physics,
University of Oxford, \\
Keble Rd., Oxford, OX1 3NP, United Kingdom}


\begin{abstract}
A systematic lowering in mass of strangeness ambiguous $pK^0_S$
peaks relative to the $nK^+$ is noted. 
We discuss how measurement of $\gamma p\to \Theta^+ K^0_S\to p K^0_S K^0_S
\to p\pi^+\pi^-\pi^+\pi^-$
in association with $\gamma p\to \Theta^+ K^0_S\to n K^+ K^0_S$ can
 help to clarify the nature of the narrow baryon state
claimed in $p K^0_S$ invariant mass distributions. This can both
establish the $\Theta^+$ in a single experiment
and also reveal, or place severe limits on, the
existence of an associated $\Sigma_5$ pentaquark.

\end{abstract}

\maketitle


\vskip 1.cm

Evidence for a pentaquark state $\Theta^+(1540)$ with positive strangeness
has been claimed in the invariant
mass of $\Theta^+\to K^+ n$ in photoproduction
~\cite{spring-8,clas,saphir,clas-2}.
Several experimental groups reported to have seen
the same narrow state in
$pK^0_S$~\cite{diana,hermes,neutrino,svd,zeus,cosy}.
The importance of devising further tests for the existence
of the $\Theta^+$ and
other predicted pentaquark baryons, $B_5$, is highlighted by concerns
that have been raised about the interpretation
of some of the current signals~\cite{rosner,dzierba,cahn}
and also the lack of a signal for such a narrow state
in other reactions, e.g.~\cite{bes,HERAB,phenix}.

For signals in $nK^+$, it is unambiguous that
any resonance $\Theta^+$ must have strangeness $S=+1$.
However, for those signals seen in $pK^0_S$, the sign of the strangeness
cannot be determined, except when an accompanying hadron
with known strangeness is also detected~\cite{diana,cosy}.
In such experiments to date the narrow state
has been assumed to be $\Theta^+$ on the grounds that ``no narrow
$\Sigma^{*+}$ is known" at such masses~\cite{hermes,neutrino,svd}.
However, one has to note that until recently
there was no evidence for a narrow $\Theta^+$ either, so
the absence of an established $\Sigma^*$ proves little
about the interpretation of such a narrow state.

\begin{figure}
\begin{center}
\epsfig{file=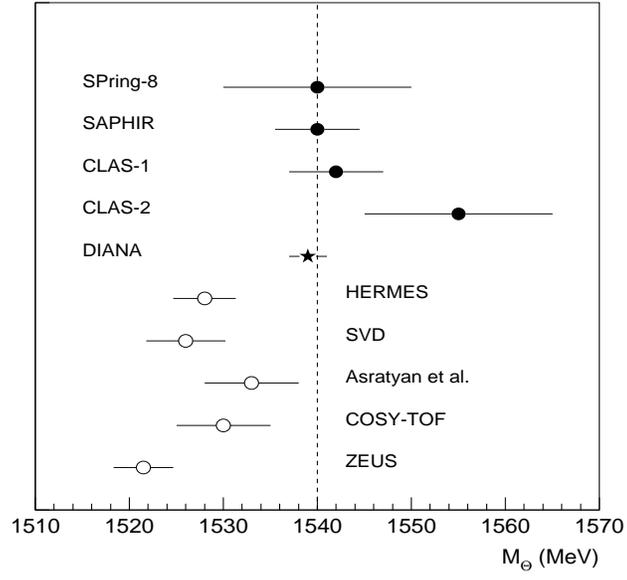, width=10cm,height=9.cm}
\caption{Experimental data for the narrow baryon displayed with
uncertainties
in its mass. The $pK^0_S$ data are shown by open circles and pure $nK^+$
by solid circles. DIANA uniquely connects $nK^+ \to pK_S^0$ and is denoted
by
a star.
}
\protect\label{fig:(1)}
\end{center}
\end{figure}

Furthermore there appears to be a systematic mass shift between signals
in $nK^+$ and $pK^0_S$ as presented in Table~\ref{tab-1} and Fig.~\ref{fig:(1)}.
It shows that the $\Theta^+$ observed in $nK^+$ invariant mass
is slightly higher than signals seen in $pK^0_S$,
except for those two~\cite{diana,cosy}, in which the strangeness
of the $pK^0_S$ is well-defined.

While the unambiguous positive strangeness of the $nK^+$ channel
makes it superficially
clean, Rosner has noted~\cite{rosner} that the diffractive background
process
$\gamma n\to n \phi\to n K^+K^-$ can be
a source of kinematic enhancement
in the $K^+ n$ channel on nuclei and thereby questions some claims
for a $\Theta^+(1540)$. Dzierba {\it et al}~\cite{dzierba} 
have shown how kinematic
reflections can even generate a narrow peak in the $K^+n$ effective mass
distribution around 1540 MeV, based on the observation that,
in addition to the $\phi$,
higher-spin mesons such as $f_2$, $a_2$, $\rho_3$ can be photoproduced and decay
to $K^+K^-$ or $K^0\overline{K^0}$. The resonance decay momenta
 for charged and neutral
$a_2$ or $\rho_3$ differ by 3-4 MeV. 
This could produce a slightly different interference pattern in $pK^0_S$ 
in contrast with that in $nK^+$, for which the interference was illustrated 
in Fig. 2 of Ref.~\cite{dzierba}.
Whether these are the source of the $\Theta^+$ peaks, and also the reason for 
the mass shift in the $nK^+$ and $pK^0_S$ enhancement, 
will require analysis of Dalitz plots in higher statistics experiments.

In this letter we advocate also studying the reaction channel
$\gamma p\to p K^0_S K^0_S$, through which any $\Theta^+$ signals
should be compared
with those reported in the $nK^+$ channel.
It is also useful for gaining information
about the pentaquark $\Sigma^+_5$. Furthermore, it is free from Rosner's
ambiguity since in
$\gamma p\to p \phi\to p K^0\overline{K^0}$
Bose symmetry allows only $K^0\overline{K^0}\to K^0_S K^0_L$; hence
 $\gamma p \to p K^0_S K^0_L$. Thus in measuring
$\gamma p \to p K^0_S K^0_S$ one can be certain that the $K^0_S$ is not the
decay of a $\phi$
and so alleviate some of the difficulty in distinguishing the source of a
measured
$K^0_S$, -- whether it comes from the $\Theta^+$ decay, or
the photo-interaction vertex~\cite{hicks}.
These remarks generalise to the higher spin mesons considered by Dzierba
 {\it et al}~\cite{dzierba}. 
The L=even/odd decays will no longer give interferences
  in the $K^0_S K^0_S$  case, thereby reducing the possibility of kinematic
  effects discussed in Ref.~\cite{dzierba}.
Therefore, measurement of $\gamma p\to \Theta^+ K^0_S\to p K^0_S K^0_S
\to p\pi^+\pi^-\pi^+\pi^-$
in association with $\gamma p\to \Theta^+ K^0_S\to n K^+ K^0_S$ can provide
a further check on $\Theta^+$ photoproduction, 
and also give evidence for or against the
existence of associated pentaquark states, $\Sigma_5$.

 In the chiral soliton model of Ref.~\cite{dpp} the $\Theta^+$
was predicted to occur in a
 $\tenbar$ flavour multiplet, while it can be assigned into a
 $\tenbar \oplus {\bf 8}_5$ in the quark model~\cite{JW,fec}.
Within both of these phenomenologies $\Sigma_5$ states 
are predicted with a mass that is
 within 100 MeV of the $\Theta^+$~\cite{JW},
and to have a photoproduction rate which is about $1/6 \sim 1/2$
that of the $\Theta^+$~\cite{cz}. 
Hence the $pK^0_S$ channel, far from being
plagued by uncertainty on whether any signal has strangeness $\pm 1$, when
compared with $nK^+$, becomes
a direct test for the pentaquark $B_5$ dynamics.

The underlying process of interest is
\be
\label{react}
\gamma p\to p K^0\overline{K^0}\to p K^0_S K^0_S\to p\pi^+\pi^-\pi^+\pi^-,
\ee
where, in principle, the $K^0_S$ events can be
rather clearly separated~\cite{saphir,hermes,svd}.
To the best of our knowledge, this channel has not been
seriously considered.
We argue that the pentaquark $\Theta^+$ signals should be compared
with those reported in the $nK^+$ channel.

In the analyses of SAPHIR~\cite{saphir}
the trigger required at least two charged particles.
From these,
kinematic reconstructions for $\gamma p\to n K^+K^0_S$
were performed in events with three charged tracks, from which
$\gamma p\to \Theta^+ K^0_S \to n K^+K^0_S$ was identified.
Since
$\gamma p \to \Theta^+ \overline{K^0}\to p K^0_S K^0_S$ should occur
{\it at the same rate} as  $\gamma p \to \Theta^+ \overline{K^0}\to n K^+
K^0_S$,
then if the $\Theta^+$ does exist, (or the $\Sigma_5$ is radically
suppressed), one would expect a narrow
state to appear in the invariant mass of $pK^0_S$ with the same mass
and with defined statistics relative to what is seen 
in $\gamma p\to nK^+ K^0_S$.
Such an observation would increase confidence in evidence for the existence
of $\Theta^+$ at 1.54 GeV with strangeness $+1$.

Possible results and their implications are:

(i) If a narrow baryon state with comparable statistics
appears in the invariant mass spectra of both $p K^0_S$ and $nK^+$,
and at the same mass,
this would support the existence of $\Theta^+$ with strangeness
$S=+1$.

(ii) If a stronger peak appears in the invariant
mass spectrum of $p K^0_S$ than in $nK^+$, and at the same mass,
it could be evidence not only for the $\Theta^+$, but also for the
production of a degenerate $\Sigma^+_5$~\cite{JW,fec,cd,cz}.
Note that the $\Sigma^+_5$
in either a $\tenbar$ or ${\bf 8}_5$ is predicted to
couple to $p K^0_S$ with a strength comparable to that
of the $\Theta^+$~\cite{cd,dpp}, while it will  
decouple from $n K^+$. 
It may also be a narrow state unless mixed with conventional $\Sigma^*$.
However, so far, all the experiments
looking for narrow states in $p K^0_S$ have claimed only
one such at around 1.53 GeV~\cite{neutrino,hermes,svd,cosy}.
As we note in Table~\ref{tab-1}, the mass of the $pK^0_S$ peak appears to be
systematically lower than that in $nK^+$. 
Though DIANA~\cite{diana} and 
COSY-TOF~\cite{cosy} determined the strangeness $+1$ of the narrow peak
in $pK^0_S$, it seems they rather confirm the systematic mass shift 
between $nK^+$ and $pK^0_S$ than solve the puzzle  
if we remember that the $NK$ scattering analyses suggest that the width of 
$\Theta$ should be narrower than 1 MeV~\cite{nk-scattering}.
Hence, 
studies with increased statistics are required to confirm or place
quantitative limits on
the production ratio of $\Sigma^+_5$ and $\Theta^+$ in the $pK^0_S$
channel~\cite{cz}.

(iii) In the case of (i), a clear $\Theta^+$ and the
absence of any other narrow structures in the region of 1.53 to 1.7 GeV
could imply either $\Sigma^+_5$ has a higher mass, or its signals
are merged into broader octet $\Sigma^*$ resonances, e.g., $\Sigma(1660)$.
However, if this indeed occurs, one at least should see
the bump structure of $\Sigma(1660)$. A further possibility is that there
is an extreme breaking of flavour symmetry such that the
$\Theta^+$ state exists but that $\Sigma_5$ does not. 
One possibility might be, if instanton forces attract
$m_{u,d} \to 0$ flavours strongly to form the $(ud)$ diquarks that help
seed the $\Theta^+$,
whereas the $m_s \neq 0$ neuters this attraction, 
the diquarks $(us)$ and $(ds)$ will be unable
to play an analogous role as the $(ud)$ in forming a $\Sigma_5$~\cite{bcd}.
Coincident with the appearance of an earlier version of the present paper 
in April 2004 (hep-ph/0404075), such ideas have been independently
developed in some detail in Ref.~\cite{vento}. 
It was shown ~\cite{vento} that
mixing between 
configurations $[ud]^0_{3_c}\bar{s}$ of Ref.~\cite{JW} and  
$([ud]^1_{6_c}\bar{s})$
of Ref.~\cite{KL} could lower the $ud\bar{s}$ energy 
to below that of $m(K) +m(u/d)$, and hence stabilise
the $(ud\bar{s})$ correlation against decay. 
This was proposed as a source of the
metastability of the $\Theta^+$. 
By contrast, the $(u\bar{d}s)$ within a $\Sigma_5$
is unstable against decay to $\pi + s$, 
whereby the dynamics of the $\Sigma_5$ could
differ significantly from those of the $\Theta^+$.

(iv) If $\Theta^+$ and $\Sigma_5^+$ are degenerate 
they may mix via the common $pK_S^0$ channel.
This would lead to two displaced eigenstates each with strangeness
components $\pm 1$.
One of these eigenstates would couple strongly to $pK_S^0$
(such as at $\sim $ 1530 MeV) and the other tend to decouple. 
However, in such a picture one might
expect both to couple to $nK^+$ through their common $S=+1$ component. Hence
it will be important
to establish if the $nK^+$ channel shows two peaks or a single broader peak
that subsumes that in the $pK_S^0$.

We note that for HERMES~\cite{hermes} 
the claimed width of 17 MeV in $pK_S^0$ appears
at variance with the rather stringent limits 
in other experiments~\cite{shanjin}.  
This could be due to degeneracy of $\Theta^+$ -$\Sigma_5^+$, 
whereby the width of
the peak is a measure of the $\Sigma_5^+$ rather than the $\Theta^+$. 
If the evidence for width $\geq 10$ MeV in $K_S^0p$ persists, then the possible
interference between a narrow $\Theta$
 and a relatively broad $\Sigma_5$ should be investigated.
 
(v) If distinct peaks are established at $\sim 1530$ MeV for $pK_S^0$ and
$\sim 1540$ MeV for
$nK^+$ then one may have to consider some rescattering
effect in the $KN$ system leading to enhancements with an electromagnetic
energy shift
between the $nK^+$ and $pK_S^0$ modes. 
The above discussion of the mechanism of Ref.~\cite{dzierba} is a 
particular example.

This work is supported,
in part, by grants from
the U.K. Engineering and Physical
Sciences Research Council (Grant No. GR/R78633/01 
and Advanced Fellowship Grant No. GR/S99433/01),
and
the Particle Physics and
Astronomy Research Council, and the
EU-TMR program ``Eurodice'', HPRN-CT-2002-00311.


\begin{table}[ht]
\begin{tabular}{c|c|c|c}
\hline
Experiments & Mass (MeV) & Width (MeV) & Observation
\\[1ex]
\hline
SPring-8~\protect\cite{spring-8} & $1540\pm 10 $ & $< 25$ & $nK^+$ \\[1ex]
\hline
SAPHIR~\protect\cite{saphir} & $1540\pm 4\pm 2$ & $<25$ & $nK^+$ \\[1ex]
\hline
CLAS-1~\protect\cite{clas} & $1542\pm 5$ & $<21$ &  $nK^+$ \\[1ex]
\hline
CLAS-2~\protect\cite{clas-2} & $1555\pm 10$ & $<26$ &  $nK^+$ \\[1ex]
\hline
DIANA~\protect\cite{diana} & $1539\pm 2 $ & $<9$ & $K^+ n\to K^0_S p$
\\[1ex]
\hline
HERMES~\protect\cite{hermes} & $1528\pm 2.6\pm 2.1$ & $17\pm 9\pm 3$ &
$pK^0_S$ \\[1ex]
\hline
SVD~\protect\cite{svd} & $1526\pm 3\pm 3 $ & $<24$ & $pK^0_S$ \\[1ex]
\hline
Asratyan {\it et al.}~\protect\cite{neutrino} & $1533\pm 5$ & $<20$ &
$pK^0_S$  \\[1ex]
\hline
ZEUS~\protect\cite{zeus} & 
$1521.5\pm 1.5\begin{array}{c} +2.8 \\ -1.7 \end{array}$ 
& $6.1\pm 1.6\begin{array}{c} +2.0 \\ -1.4 \end{array}$ 
& $ p K^0_S$, $\bar{p}K^0_S$ \\[1ex]
\hline
COSY-TOF~\protect\cite{cosy} & $1530\pm 5$ & $<18\pm 4$ & $ pp\to \Sigma^+ p
K^0_S$\\
[1ex]
\hline
\end{tabular}
\caption{Experimental signals for the narrow baryon observed in the
invariant mass
of $nK^+$ and $pK^0_S$. Note that the strangeness ambiguous $pK^0_S$ signals
appear to be systematically
lower than those in the $nK^+$.}
\label{tab-1}
\end{table}


\end{document}